\documentclass[] {iopart}

\usepackage{graphicx}
\usepackage{dcolumn}
\usepackage{bm}
\usepackage{citesort}
\begin{document}

\title{High temperature ferromagnetism in Co-implanted TiO$_2$ rutile}

\author{Numan Akdogan$^1,\footnote{Author to whom correspondence should be addressed.}$, Alexei Nefedov$^2$, Hartmut Zabel$^3$,
Kurt Westerholt$^3$, Hans-Werner Becker$^4$, Christoph Somsen$^5$,
\c{S}afak G\"{o}k$^6$, Asif Bashir$^7$, Rustam Khaibullin$^{8,9}$
and Lenar Tagirov$^{8,9}$}

\address{$^1$ Department of Physics, Gebze Institute of Technology, 41400 Kocaeli, Turkey}
\address{$^2$ Lehrstuhl f\"{u}r Physikalische Chemie I, Ruhr-Universit\"{a}t Bochum, D-44780 Bochum, Germany}
\address{$^3$ Institut f\"{u}r Experimentalphysik/Festk\"{o}rperphysik, Ruhr-Universit\"{a}t Bochum, D-44780 Bochum, Germany}
\address{$^4$ Institut f\"{u}r Physik mit Ionenstrahlen, Ruhr-Universit\"{a}t Bochum, D-44780 Bochum, Germany}
\address{$^5$ Institut f\"{u}r Werkstoffe, Ruhr-Universit\"{a}t Bochum, D-44780 Bochum, Germany}
\address{$^6$ Lehrstuhl f\"{u}r Angewandte Festk\"{o}rperphysik, Ruhr-Universit\"{a}t Bochum, D-44780 Bochum, Germany}
\address{$^7$ Lehrstuhl f\"{u}r Physikalische Chemie I, Ruhr-Universit\"{a}t Bochum, D-44780 Bochum, Germany}
\address{$^8$ Kazan Physical-Technical Institute of RAS, 420029 Kazan, Russia}
\address{$^9$ Kazan State University, 420008 Kazan, Russia}

\ead{numan.akdogan@ruhr-uni-bochum.de}

\begin{abstract}
We report on structural, magnetic and electronic properties of
Co-implanted TiO$_2$ rutile single crystals for different
implantation doses. Strong ferromagnetism at room temperature and
above is observed in TiO$_2$ rutile plates after cobalt ion
implantation, with magnetic parameters depending on the cobalt
implantation dose. While the structural data indicate the presence
of metallic cobalt clusters, the multiplet structure of the Co
\emph{L}$_3$ edge in the XAS spectra gives clear evidence for a
substitutional Co$^{2+}$ state. The detailed analysis of the
structural and magnetic properties indicates that there are two
magnetic phases in Co-implanted TiO$_2$ plates. One is a
ferromagnetic phase due to the formation of long range ferromagnetic
ordering between implanted magnetic cobalt ions in the rutile phase,
and the second one is a superparamagnetic phase originates from the
formation of metallic cobalt clusters in the implanted region. Using
x-ray resonant magnetic scattering, the element specific
magnetization of cobalt, oxygen and titanium in Co-implanted TiO$_2$
single crystals are investigated. Magnetic dichroism was observed at
the Co $\emph{L}_{2,3}$ edges as well as at the O \emph{K} edge. The
interaction mechanism, which leads to ferromagnetic ordering of
substituted cobalt ions in the host matrix, is also discussed.
\end{abstract}


\submitto{\JPD}

\section{Introduction}

Oxide-based diluted magnetic semiconductors (DMSs) have recently
attracted considerable attention because of reports on the room
temperature ferromagnetism (FM) in several systems and their
projected potential for spintronic devices
\cite{MatsukuraElsevier02,JanischJPCM05}. Since Matsumoto \emph{et
al.} \cite{MatsumotoSci01} observed room temperature FM in Co-doped
anatase TiO$_2$, much interest has been focused on the titanium
dioxide as a host material for magnetic doping. Co-doped TiO$_2$ has
been grown by using a wide variety of growth methods, including
pulsed laser deposition (PLD)
\cite{KimAPL02,StampeJAP03,ShindePRB03,HigginsPRB04}, laser
molecular beam epitaxy (LMBE)
\cite{KimPRL03,ToyosakiNature04,MurakamiJAP04}, combinatorial LMBE
\cite{MatsumotoSci01,FukumuraJJAP03}, reactive co-sputtering
\cite{ParkJAP02,RameevJMMM03}, magnetron sputtering
\cite{YangJAP04,BalagurovJETP04}, metal organic chemical-vapor
deposition (MOCVD) \cite{SeongAPL02}, oxygen plasma assisted
molecular beam epitaxy (OPA-MBE)
\cite{ChambersAPL01,ChambersTSF02,ChambersAPL03} and as well as the
sol-gel method \cite{SooAPL02}. Both the epitaxial TiO$_2$ anatase
thin film and the single-crystalline TiO$_2$ rutile have also been
doped by using ion implantation technique
\cite{KimAPL03,KhaibullinJPCM04,AkdoganJPCM05,AkdoganJMMM06,NefedovAPL06,PintoEPJB07,AkdoganSM07}.
In addition to different growth techniques, different substrates
such as Al$_2$O$_3$ \cite{MatsumotoJJAP01,SuryanarayananSSC05},
SrTiO$_3$
\cite{ChambersAPL01,KimAPL02,ChambersTSF02,ShindePRB03,StampeJAP03,ChambersAPL03,BalagurovJETP04},
LaAlO$_3$
\cite{MatsumotoSci01,ChambersTSF02,KimPRL03,ShindePRB03,StampeJAP03},
Si \cite{ParkJAP02} and SiO$_2$/Si \cite{SeongAPL02} have been used
to synthesize Co-doped TiO$_2$ films.

Many groups have observed room temperature ferromagnetism in
Co-doped TiO$_2$ for both anatase and rutile phases
\cite{MatsumotoSci01,ParkJAP02,ShindePRB03,RameevJMMM03,PrellierJPCM03,KhaibullinJPCM04,AkdoganJPCM05,JanischJPCM05,AkdoganJMMM06,NefedovAPL06,AkdoganSM07,KhaibullinNIMB07}.
A Curie temperature of about 650 K \cite{ShindePRB03} and 700 K
\cite{KhaibullinJPCM04} was reported by different groups. Subsequent
reports have concentrated on the origin of ferromagnetism in this
material. Spectroscopic studies indicated that cobalt ions in thin
TiO$_2$ films exist in a +2 formal oxidation state, consistent with
ferromagnetism originating from Co substitution on the Ti site
\cite{KimAPL02}. In other publications it is suggested that the
ferromagnetic behavior is due to cobalt clustering depending on the
growth conditions \cite{MatsumotoSci01,ChambersTSF02,KimPRL03}.
Chambers \emph{et al.} \cite{ChambersAPL01} reported that the
solution of Co in TiO$_2$ is possible at least up to 10\% when the
TiO$_2$ is deposited on SrTiO$_3$ substrate. However, when the
TiO$_2$ films grown on LaAlO$_3$ substrate the solid solution is
about 2-7\% \cite{MatsumotoSci01,ShindePRB03}. Co metal clusters
were observed in the as-grown Co-doped TiO$_2$ films with a cobalt
concentration of 2\%. Post-annealing of the samples leads to
dissolving of clusters in the TiO$_2$ matrix \cite{ShindePRB03}. For
higher cobalt concentrations, bigger cobalt cluster were reported
with a cluster size of about 150 nm \cite{SeongAPL02}.

If the observed ferromagnetism is actually due to substituted
magnetic elements in the host matrix, then another important
question arises; what is the coupling mechanism which leads to
ferromagnetism? Recently, we have reported the observation of room
temperature FM and in-plane magnetic anisotropy of
single-crystalline TiO$_2$ rutile structures after high dose Co
implantation \cite{AkdoganJPCM05,AkdoganJMMM06,AkdoganSM07}. From
the observation of the in-plane magnetic anisotropy we concluded
that FM in this system results from the incorporation of Co ions in
the TiO$_2$ lattice, but a co-existence with Co nanoclusters could
not be excluded.

In order to clarify this situation we studied the structural,
magnetic and electronic properties of Co-doped (100)-oriented rutile
TiO$_2$ single crystals for different implantation doses. The
resulting Co:TiO$_2$ samples have been characterized by Rutherford
backscattering spectroscopy (RBS) to obtain the Co depth
distribution profiles and by atomic force microscopy (AFM) to check
the surface properties after implantation, as well as by x-ray
diffraction (XRD) and by high resolution transmission electron
microscopy (TEM) to reveal the presence of precipitates and metallic
Co clusters. X-ray absorption spectroscopy (XAS) has also been
employed to determine whether the implanted cobalt ions are in the
Co$^{2+}$ oxidation state or are in the metallic state. The magnetic
properties of TiO$_2$ rutile samples have been investigated using
magneto-optical Kerr effect (MOKE), superconducting quantum
interference device (SQUID) based magnetometry, and x-ray resonant
magnetic scattering (XRMS) techniques. In addition, Hall effect
measurements were carried out to verify the occurrence of intrinsic
ferromagnetism and relate it to the carrier type in the samples.

\section{Sample Preparation}

40 keV Co$^+$ ions were implanted into (100)-oriented
$15\times15\times1$ mm$^{3}$ single-crystalline TiO$_2$ rutile
substrates (from Moscow Power Engineering Institute in Russia) by
using the ILU-3 ion accelerator (Kazan Physical-Technical Institute
of the Russian Academy of Science) with an ion current density of $8
\mu A\cdot cm^{-2}$. The implantation dose varied in the range of
$0.25-1.50\times10^{17} ions\cdot cm^{-2}$. The sample holder was
cooled by flowing water during the implantation to prevent the
samples from overheating. The implanted plates were cut by a diamond
cutter into smaller pieces for structural, magnetic and electronic
studies. As a last step, four gold contacts were evaporated on the
corners of the samples for Hall effect measurements. The list of the
Co-implanted TiO$_2$ samples used in the present study is given in
Table~\ref{table1}.

\begin{table}[!h]
\caption{\label{table1}TiO$_2$ samples implanted with 40 keV Co$^+$
for different Co ion doses.}
\begin{indented}
\item[]\begin{tabular}{cc}
\hline
Sample & Dose ($\times10^{17} ion\cdot cm^{-2}$) \\
\hline
1 & 0.25 \\
2 & 0.50 \\
3 & 0.75 \\
4 & 1.00 \\
5 & 1.25 \\
6 & 1.50 \\
\hline
\end{tabular}
\end{indented}
\end{table}

\section{Experimental Results}

\subsection{Structural Properties}

\begin{figure}[!h]
\centering
\includegraphics{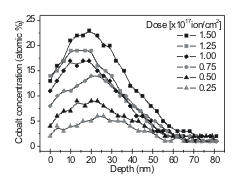}
\caption{\label{figure1} The cobalt concentration profile as a
function of depth and for different implantation doses measured by
RBS.}
\end{figure}

In this section, the structural properties of non-implanted and
Co-implanted TiO$_2$ rutile plates are presented. The depth
distribution of implanted cobalt ions in the rutile samples as well
as the cobalt concentration for each dose are determined by using
the RBS technique. The RBS measurements were carried out at the
Dynamic Tandem Laboratory (DTL) at the Ruhr-Universit\"{a}t Bochum.
Fig.~\ref{figure1} presents the depth dependence of the cobalt
concentration in Co-implanted TiO$_2$ plates for different Co ion
implantation doses. The RBS data show a maximum cobalt concentration
of about 25 at. \% for the highest Co dose and it decreases to about
5 at. \% for the lowest dose. Due to the ion sputtering of the
surface during implantation, the maximum slightly shifts to the left
for higher dose levels. An extended inward tail up to 70 nm due to
cobalt diffusion into the volume of the rutile single crystals is
also observed for each implantation dose.

\begin{figure}[!h]
\centering
\includegraphics{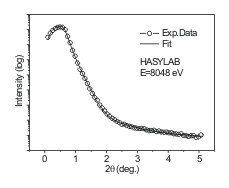}
\caption{\label{figure2} Small angle x-ray reflectivity data and fit
for sample 6 ($1.50\times10^{17} ions\cdot cm^{-2}$).}
\end{figure}

\begin{figure}[!h]
\centering
\includegraphics{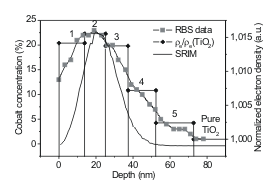}
\caption{\label{figure3} The cobalt concentration (RBS data) and the
electron density determined from the small angel x-ray reflectivity
as a function of depth for sample 6 ($1.50\times10^{17} ions\cdot
cm^{-2}$). The solid line represents the calculated SRIM profile.}
\end{figure}

Fig.~\ref{figure2} shows small-angle x-ray reflectivity taken with
synchrotron radiation at the "Hamburg Synchrotron Radiation
Laboratory" (HASYLAB) with an energy of E=8048 eV. The solid line in
Fig.~\ref{figure2} is a fit to the data points for sample 6
($1.50\times10^{17} ions\cdot cm^{-2}$) from Table~\ref{table2} as
is obtained by the commercial software WinGIXA, which is based on
the Parratt formalism \cite{ParrattPR54}. Since the cobalt
concentration in the TiO$_2$ crystals changes with the depth, for
fitting of the reflectivity data the implanted area is sliced into
five layers. The roughness and electron density values obtained from
the fit are listed in Table~\ref{table2} for each layer. The model
used for fitting perfectly matches the RBS data and
Fig.~\ref{figure3} shows the depth dependence of the cobalt
concentration and the normalized electron density
($\rho_e/\rho_e$(TiO$_2$)) obtained from the fit of reflectivity.
The solid line in Fig.~\ref{figure3} presents the calculated profile
using the SRIM (\textit{Stopping and Range of Ions in Matter})
software \cite{ZieglerPP85}, without taking ion sputtering effects
into account.

\begin{table}[!h]
\caption{\label{table2}Fitting parameters of reflectivity curve for
sample 6 ($1.50\times10^{17} ions\cdot cm^{-2}$).}
\begin{indented}
\item[]\begin{tabular}{cccc}
\hline
Layer & Thickness (nm) & Roughness (~\AA) & $\rho_e$ ($g\cdot cm^{-3}$)  \\
\hline
1. layer & 11.2 & 26.87 & 4.229 \\
2. layer & 11.1 & 0.50 & 4.234\\
3. layer & 12.1 & 0.71 & 4.227 \\
4. layer & 14.2 & 0.09 & 4.200 \\
5. layer & 20.3 & 9.49 & 4.180 \\
Pure TiO$_2$ & --- & 0.09 & 4.170 \\
\hline
\end{tabular}
\end{indented}
\end{table}

\begin{figure}[!h]
\centering
\includegraphics{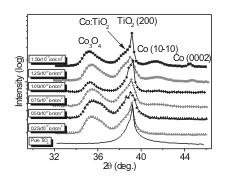}
\caption{\label{figure4} High-angle Bragg scattering scan for
non-implanted (solid line) and different dose implanted
(100)-TiO$_2$ samples. The presence of Co clusters is clearly seen
for the highest dose (Sample 6).}
\end{figure}

\begin{figure}[!h]
\centering
\includegraphics{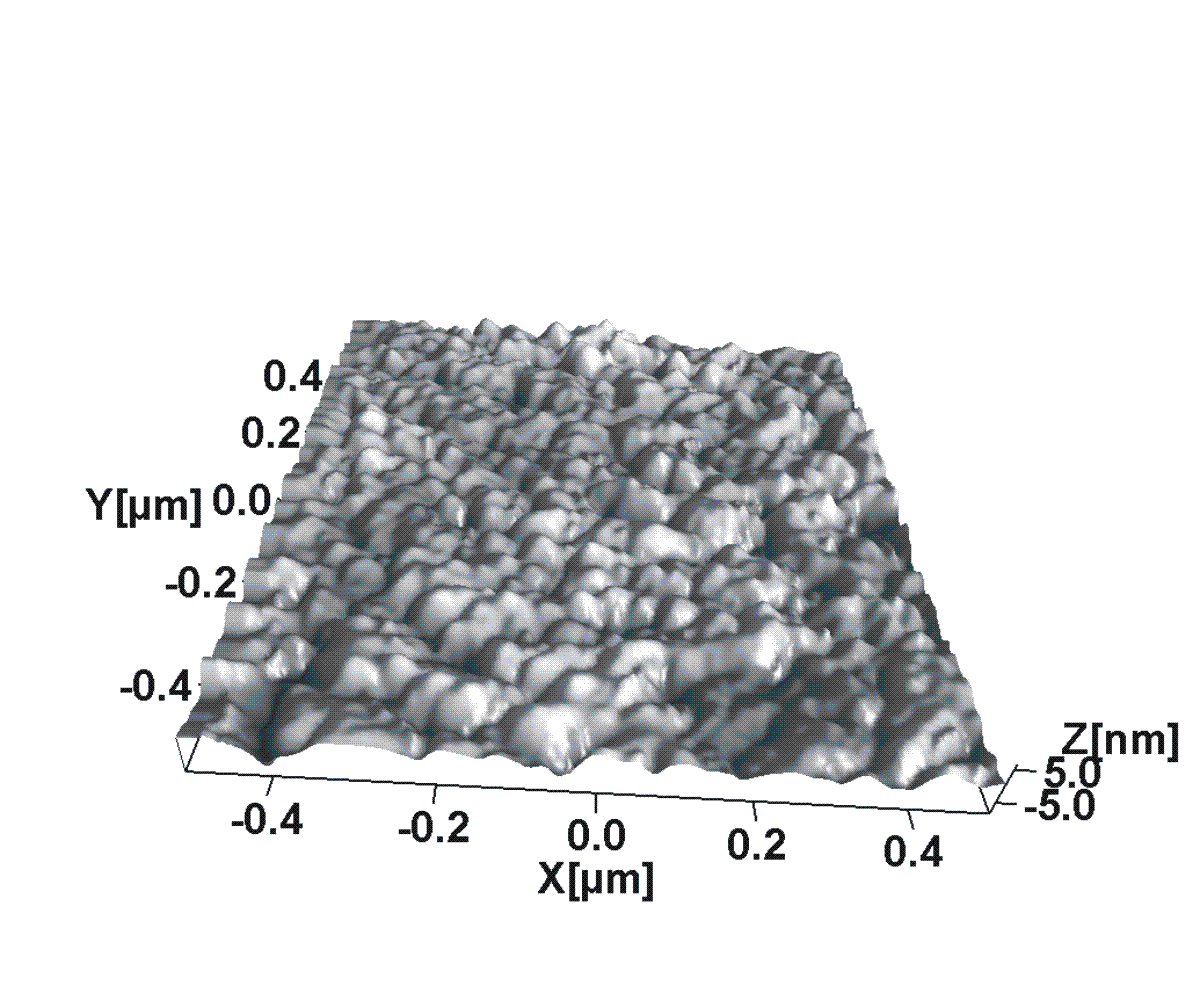}
\caption{\label{figure5} AFM surface topography of (100)-TiO$_2$
rutile after Co ion implantation with a dose of $1.25\times10^{17}
ions\cdot cm^{-2}$ (Sample 5).}
\end{figure}

The high angle XRD measurements were also carried out at HASYLAB, in
order to detect possible impurity phases in the samples after
implantation. The Bragg scans before and after implantation with
different doses are shown in Fig.~\ref{figure4} for (100)-oriented
TiO$_2$ rutile samples. Increase of the implantation dose up to
$1.50\times10^{17} ions\cdot cm^{-2}$ results in two additional
peaks which correspond to the ($10\overline{1}0$) and (0002)
reflections of hcp Co. Below the implantation dose of
$1.25\times10^{17} ions\cdot cm^{-2}$ cobalt nanoclusters cannot be
detected by x-ray diffraction. For every implanted sample a tail,
indicated by an arrow in Fig.~\ref{figure4}, is present around the
main peak of the TiO$_2$ (200) reflection. This tail results from
the expansion of the TiO$_2$ lattice upon cobalt implantation and is
not observed before the implantation. In addition, a new peak is
present on the low angle side which corresponds to the spinel cobalt
oxide (Co$_3$O$_4$) phase reported already by Khaibullin \emph{et
al.} in Co-implanted TiO$_2$ \cite{KhaibullinNIMB07}. Due to the
difference in etching rates of Co and the TiO$_2$ during high dose
ion implantation \cite{KhaibullinNIMB07}, cobalt nanoparticles form
on the surface and become oxidized forming antiferromagnetic
Co$_3$O$_4$ with a Neel temperature of about 40 K.
Fig.~\ref{figure5} presents the surface morphology of sample 5
($1.25\times10^{17} ions\cdot cm^{-2}$) probed by AFM (Digital
Instruments NanoScope MultiMode AFM). The AFM image clearly shows a
network of cobalt oxide islands on the surface with a roughness of
about $2.14\pm0.25$ nm.

\begin{figure}[!h]
\centering
\includegraphics[width=0.5\textwidth]{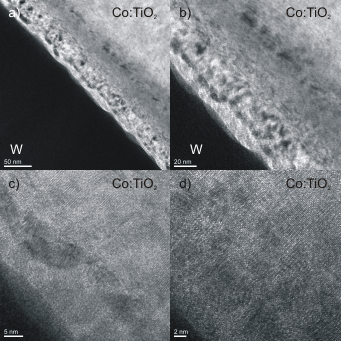}
\caption{\label{figure6} Cross-sectional TEM images of sample 6
($1.50\times10^{17} ions\cdot cm^{-2}$).}
\end{figure}

For further investigations on the effects of ion implantation into
TiO$_2$, high resolution cross sectional TEM measurements were
performed. For the preparation of TEM samples, the plates were
thinned by focused ion beam (FIB) technique. First, the sample
surface is covered by a tungsten (W) film to prevent charging
effects. Then a very small cross sectional piece of the implanted
sample was cut by FIB. Fig.~\ref{figure6} presents TEM images of
sample 6 ($1.50\times10^{17} ions\cdot cm^{-2}$) with an increasing
resolution from 50 nm to 2 nm. In Figs.~\ref{figure6}(a) and (b), a
general overview of the sample is shown. It can clearly be seen that
a surface layer of about 40 nm thickness is strongly damaged after
ion bombarding. There are many defects and differently sized cobalt
clusters in this region. However, in Figs.~\ref{figure6}(c) and (d)
it can be recognized that the structure of TiO$_2$ is preserved
after implantation. Beneath the surface layer there is another
cobalt rich layer of about 40 nm thickness. Element specific TEM
measurements indicate that the cobalt concentration in this layer is
much smaller than in the surface layer in agreement with the RBS and
x-ray reflectivity data (Fig.~\ref{figure3}).

\subsection{Magnetic Properties}

\subsubsection{{\bf In-plane magnetic anisotropies and hysteresis measurements}}

\begin{figure}[!h]
\centering
\includegraphics[width=0.75\textwidth]{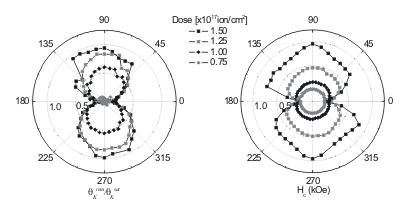}
\caption{\label{figure7} Azimuthal dependence of the normalized
remanent magnetization (left) and the coercive field (right) for
different Co ion doses.}
\end{figure}

In order to investigate the in-plane magnetic anisotropy of the
implanted samples we used a high-resolution MOKE setup in the
longitudinal configuration with s-polarized light
\cite{ZeidlerPRB96,SchmitteJAP02,WestphalenRSI07}. The MOKE setup
allows for a rotation of the sample around its surface normal (by
the angle $\varphi$) in order to apply a magnetic field in various
in-plane directions and thus provide information about the in-plane
magnetic anisotropy. The in-plane magnetic anisotropy of the samples
doped with different doses as determined by the MOKE measurements
are shown in Fig.~\ref{figure7}. For the sample with the highest
dose, both, the remanent Kerr signal normalized to the Kerr signal
at saturation ($\theta_{K}^{rem}/\theta_{K}^{sat}$) and coercive
field ($H_C$), are reduced to almost zero near the hard axis
($\varphi=0^\circ-180^\circ$), while for the magnetic field applied
along the easy axis ($\varphi=90^\circ-270^\circ$) they are close to
unity. It is evident from Fig.~\ref{figure7} that both
$\theta_{K}^{rem}/\theta_{K}^{sat}$ and $H_C$ exhibit a strong
two-fold symmetry for the highest dose which decreases with
decreasing implantation dose.

\begin{figure}[!h]
\centering
\includegraphics{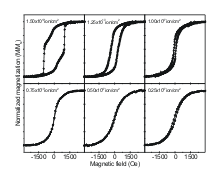}
\caption{\label{figure8} SQUID hysteresis loops for different Co ion
doses taken parallel to the easy axis at T=300 K.}
\end{figure}

\begin{figure}[!h]
\centering
\includegraphics{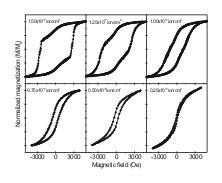}
\caption{\label{figure9} SQUID hysteresis loops for Co-implanted
TiO$_2$ samples measured at 5 K along the easy axis.}
\end{figure}

Hysteresis loops of the Co-implanted TiO$_2$ samples, obtained by
using a Quantum Design MPMS XL SQUID magnetometer, are presented in
Figs.~\ref{figure8} and ~\ref{figure9}. Fig.~\ref{figure8} shows
hysteresis curves taken parallel to the easy axis at 300 K. For the
highest dose ($1.50\times10^{17} ions\cdot cm^{-2}$) a square-like
hysteresis curve is observed with a large coercive field of
$H_C$=950 Oe. A rather sharp magnetization reversal takes place for
this sample with a small step at 260 Oe. For the samples implanted
with intermediate ion doses ($1.00-1.25\times10^{17} ions\cdot
cm^{-2}$), the recorded $M-H$ loops also show hysteretic behavior,
but the coercive fields decrease significantly. The remanent
magnetization normalized to the saturation magnetization also
decreases for the intermediate doses. At 5 K the two step feature in
the hysteresis curve is not only present for the highest dose but
also for intermediate doses (Fig.~\ref{figure9}). The low dose
implanted samples exhibit at low temperatures a typical
superparamagnetic behavior with a pronounced paramagnetic
contribution to the hysteresis curves .

\begin{figure}[!h]
\centering
\includegraphics[width=0.5\textwidth]{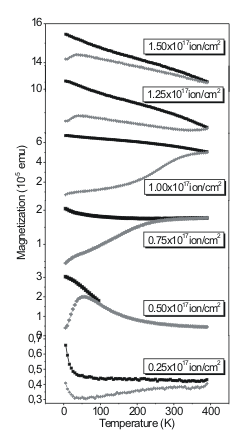}
\caption{\label{figure10} FC (black squares) and ZFC (grey circles)
magnetization curves of Co-implanted TiO$_2$ rutile samples taken
using a SQUID magnetometry.}
\end{figure}

In order to further investigate the effect of cobalt clusters, we
have performed temperature dependent magnetization ($M-T$)
measurements using a SQUID magnetometer. Fig.~\ref{figure10}
presents field cooled (FC) and zero field cooled (ZFC) plots for
each sample. For ZFC measurements, the samples are cooled in zero
field to 5 K and the magnetization is recorded during warming up to
390 K in an applied field of 100 Oe parallel to the film surface.
For FC measurements the applied field of 100 Oe is kept during
cooling to 5 K and the magnetization is recorded during field
warming with the same field value. The FC (black squares) and ZFC
(grey circles) curves diverge substantially for all doses and the
peak in the ZFC curve progressively shifts to higher temperatures
with increasing cobalt concentration. This behavior is not expected
for a ferromagnet and suggests the presence of magnetic cobalt
nanoparticles in the films or a spin-glass like nature of the system
\cite{ShandPRB98,HochepiedJAP00}. The $M-T$ curve of sample 1
($0.25\times10^{17} ions\cdot cm^{-2}$) is rather unusual and may be
attributed to the coexistence of a weak ferromagnetic and a
superparamagnetic phase with a transition temperature of about 30 K.
The $M-T$ curves for films with higher cobalt concentrations
($0.50-0.75\times10^{17} ions\cdot cm^{-2}$) indicate the occurrence
of superparamagnetism with a blocking temperature of about 100 K and
250 K for sample 2 ($0.50\times10^{17} ions\cdot cm^{-2}$) and
sample 3 ($0.75\times10^{17} ions\cdot cm^{-2}$), respectively. The
temperature dependence of sample 4 is similar to that of sample 2
and sample 3 except the blocking temperature is much higher, namely
above 390 K. It is also important to note that the FC curve of this
sample shows a more or less continuous behavior versus temperature
which is typical for ferromagnets. The reported room temperature
ferromagnetism with a two fold in-plane magnetic anisotropy in this
sample indicates that for this dose substituted cobalt ions start to
interact ferromagnetically. The FC and ZFC curves of sample 5
($1.25\times10^{17} ions\cdot cm^{-2}$) and sample 6
($1.50\times10^{17} ions\cdot cm^{-2}$) are much closer to each
other. This progression indicates that the ferromagnetic phase
becomes dominant in these samples. The observation of a two
component hysteresis at RT for sample 6 supports this argument.
Small peaks in the ZFC curves at low temperatures of these samples
indicate the existence of superparamagnetic cobalt clusters. These
clusters are also clearly seen in the TEM images of sample 6
(Fig.~\ref{figure6}).

\subsubsection{{\bf XRMS and XAS measurements}}

To shed more light on the origin of room temperature ferromagnetism
in Co-doped TiO$_2$, the magnetic properties of Co-implanted TiO$_2$
rutile films have also been investigated using the XRMS and XAS
techniques. Both the XRMS and XAS experiments were carried out at
the undulator beam lines UE56/1-PGM and UE52-SGM at BESSY II
(Berlin, Germany) using the ALICE diffractometer \cite{GrabisRSI03}.
The diffractometer comprises a two circle goniometer and works in
horizontal scattering geometry. A maximum magnetic field of
$\pm2700Oe$ can be applied in the scattering plane along the sample
surface either parallel or antiparallel to the photon helicity,
which corresponds to the longitudinal magneto-optical Kerr effect
(L-MOKE) geometry. The magnetic contribution to the scattered
intensity (XRMS) was always measured by reversing the magnetic field
direction while keeping the photon helicity fixed. Thus, by tuning
the energy to the Co $L_3$ absorption edge (780 eV), reflectivity
scans were taken and the magnetic splitting for plus and minus field
was clearly seen (presented in Ref. \cite{NefedovAPL06}). As a
compromise between high scattering intensity and high magnetic
sensitivity for the investigation of the magnetic properties via
energy scans at the Co \emph{L} edges, the scattering angle was
fixed at the position of $2\theta=8.2^\circ$ (the angle of incidence
$\theta=4.1^\circ$) \cite{NefedovAPL06}. For measurements at the O
\emph{K} edge (E=535 eV) the scattering angle was fixed at
$2\theta=12^\circ$, which corresponds to the same scattering vector
in the reciprocal space.

\begin{figure}[!h]
\centering
\includegraphics{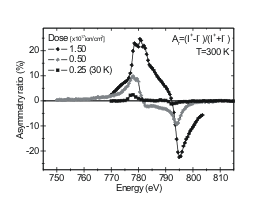}
\caption{\label{figure11} Dose dependence of the asymmetry ratio at
the Co $\emph{L}_{2,3}$ edges measured at saturation field.}
\end{figure}

\begin{figure}[!h]
\centering
\includegraphics{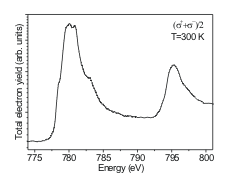}
\caption{\label{figure12} X-ray absorption spectra of sample 6
measured at the Co $L_{2,3}$ edges, determined by total electron
yield. $\sigma^+$ and $\sigma^-$ denote the right and left circular
polarization of the incident light, respectively.}
\end{figure}

First, we measured the energy dependence of the scattered intensity
(XRMS) around the Co $\emph{L}_{2,3}$ edges. Since the magnetic
contribution to the resonant scattering can best be visualized by
plotting the asymmetry ratio ($A_r=(I^+-I^-)/(I^++I^-)$), in
Fig.~\ref{figure11} we present the dose dependence of the asymmetry
ratio at the Co \emph{L} edges measured in saturation at room
temperature. Only the lowest dose of $0.25\times10^{17} ions\cdot
cm^{-2}$ is measured at 30 K. The magnetization of the samples
decreases by decreasing the Co ion dose in agreement with SQUID
hysteresis curves and previous MOKE measurements \cite{AkdoganSM07}.
It is important to note that the fine structure around the Co
$\emph{L}_3$ edge, which is clearly seen in the asymmetry ratio in
Fig.~\ref{figure11} for sample 6, is not typical for metallic
cobalt. It is well known that in the case of metallic films the
absorption spectra around the $\emph{L}_3$ peak of Co consists of a
single component \cite{ReganPRB01}. This fine structure of the Co
$\emph{L}_3$ peak is similar to that observed before for CoO films
\cite{ReganPRB01}, and it is indicative of a Co$^{2+}$ state. To
make this situation more clear and to record comparable results with
previous reports \cite{KimPRL03}, we performed XAS experiments. The
absorption data were taken by the total electron yield (TEY) method,
i.e. by measuring the sample drain current. Since the external
magnetic field changes the excited electron trajectories, the XAS
spectra were measured with fixed photon helicity in remanence. The
angle of incidence was set again to $4.1^\circ$ with respect to the
surface. The absorption spectra were normalized to the incoming
photon flux measured from the beam line mirror. The averaged x-ray
absorption spectra ($\sigma^+$+$\sigma^-$)/2 at the Co $L_{2,3}$
edges is shown in Fig.~\ref{figure12}. The XAS spectrum clearly
shows a multiplet structure at the Co $L_3$ edge. This multiplet
structure is a strong indication for the Co ions being in the
Co$^{2+}$ state in this sample.

\begin{figure}[!h]
\centering
\includegraphics{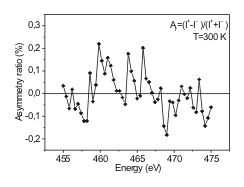}
\caption{\label{figure13} Asymmetry ratio measured at the Ti
$\emph{L}_{2,3}$ edges for sample 6.}
\end{figure}

\begin{figure}[!h]
\centering
\includegraphics{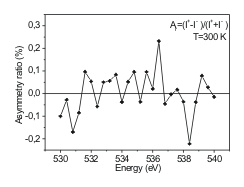}
\caption{\label{figure14} Asymmetry ratio taken at the O \emph{K}
absorption edge for sample 6.}
\end{figure}

The magnetic signal at the Ti $\emph{L}_{2,3}$ and the O \emph{K}
edges was also investigated for sample 6. Fig.~\ref{figure13} and
Fig.~\ref{figure14} show the corresponding asymmetry ratios. Within
the sensitivity limit no magnetic signal could be recorded for Ti
and O at room temperature. However, at the O \emph{K} edge, a small
but clearly visible magnetic signal was observed at T=30 K
\cite{NefedovAPL06}. It should be mentioned that the oxygen
polarization has also been observed for samples with lower dose
(dose levels of $1.00-1.25\times10^{17} ions\cdot cm^{-2}$). For
samples implanted with doses below $1.00\times10^{17} ions\cdot
cm^{-2}$, the magnetic signal at the O \emph{K} edge is below the
sensitivity limit of the experimental setup.

\subsubsection{{\bf High temperature magnetization experiments}}

\begin{figure}[!h]
\centering
\includegraphics{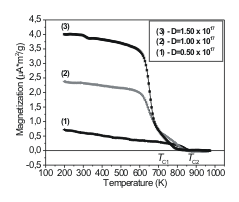}
\caption{\label{figure15} High temperature magnetization curves
measured in a field of 2000 Oe for TiO$_2$ rutile samples implanted
by cobalt ions with different doses.}
\end{figure}

In order to determine the Curie temperatures ($T_C$) of Co:TiO$_2$
samples, we have carried out thermo-magnetic measurements using the
Faraday balance technique \cite{BourovAG96} by heating the samples
from 100 K (ZFC) up to 1000 K with a rate of 100 K/min in air and at
an applied field of 2 kOe. In Fig.~\ref{figure15} we show the
magnetization curves for TiO$_2$ rutile plates implanted with
different doses. The sample with a dose of $0.50\times10^{17}
ions\cdot cm^{-2}$ shows the magnetic/non-magnetic transition
temperature at about 850 K (curve 1). For the samples 4
($1.00\times10^{17} ions\cdot cm^{-2}$) and 6 ($1.50\times10^{17}
ions\cdot cm^{-2}$), two magnetic ordering temperatures of
$T_{C1}$$\sim$700 K and $T_{C2}$$\sim$850 K, were observed (curves 2
and 3). This shows that two ferromagnetic phases, a ``low
temperature'' and a high-temperature'' phase, coexist in these
samples. The contribution to the magnetization from the
high-temperature phase decreases gradually with increasing the
cobalt implantation dose. Finally, for the sample with the highest
dose of $1.50\times10^{17} ions\cdot cm^{-2}$, the low-temperature
phase dominates, while the high-temperature phase practically
disappears (curve 3).

It should be stated here that the high temperature magnetization
curves presented in Fig.~\ref{figure15} are irreversible, i.e. on
cooling down the ferromagnetic signal disappears. From this we infer
that some diffusion process or recrystallization may occur, which
should not be confused with a real Curie temperature. Furthermore,
the magnetization versus temperature does not follow the shape of a
usual order parameter. On the other hand, after vacuum annealing the
samples at high temperatures \cite{KhaibullinNIMB07}, $T_{C1}$
reappears but not $T_{C2}$, indicating that $T_{C1}$ is more
intrinsic than $T_{C2}$. This point clearly needs some further
investigations. For the present purpose, it is important to note
that a stable ferromagnetic phase exists at room temperature and far
beyond, which may be very useful for high temperature applications
of DMSs.

\section{Hall effect measurements}

The observation of an anomalous Hall effect (AHE) is suggested to be
one of the important criteria for DMS materials to be intrinsically
ferromagnetic \cite{OhnoAPL96,MatsukuraPRB98}. In the past, several
groups reported the AHE in highly reduced TiO$_2$ films doped with
either with Co or Fe, from which they infer the possibility of
intrinsic ferromagnetism in these samples
\cite{WangAPL03,ToyosakiNature04,HigginsPRB04}. However, recently,
Shinde \emph{et al.} \cite{ShindePRL04} reported the co-occurrence
of superparamagnetism and AHE in highly reduced Co-doped TiO$_2$
rutile films, raising questions about the usefulness of the AHE as a
test of the intrinsic nature of ferromagnetism in DMS materials
without a detailed characterization of the sample.

In magnetic materials, in addition to the ordinary Hall effect
(OHE), there is an additional voltage proportional to the sample
magnetization \cite{HurdPP72}, the so-called anomalous Hall effect.
Hence, the Hall voltage can be written as follows \cite{HurdPP72},

\begin{center}
\begin{equation}\label{Hallvoltage}
V_H = \Big(\frac{R_0I}{t}\Big)Hcos\alpha +
\Big(\frac{R_A\mu_0I}{t}\Big)Mcos\theta,
\end{equation}
\end{center}

where $t$ is the film thickness and $I$ is the current. $R_0$ and
$R_A$ are the ordinary and anomalous Hall effect coefficients,
respectively. $\mu_0$ is the permeability of free space. $\alpha$ is
the angle between the applied magnetic field ($H$) and the film
normal. $\theta$ is the angle between the sample magnetization ($M$)
and the sample normal. The first term in Eq. \ref{Hallvoltage} is
the ordinary Hall effect and arises from the Lorentz force acting on
conduction electrons. This establishes an electric field
perpendicular to the applied magnetic field and to the current. The
second term is the anomalous Hall effect and it is conventionally
attributed to spin dependent scattering mechanism involving a
spin-orbit interaction between the conduction electrons and the
magnetic moments of the material. At low applied magnetic fields,
the Hall voltage ($V_H$) is dominated by the magnetic field
dependence of the sample magnetization $M$. When the applied
magnetic field is high enough to saturate the sample magnetization,
the magnetic field dependence of the Hall voltage becomes linear due
to the ordinary Hall effect.

\begin{figure}[!h]
\centering
\includegraphics{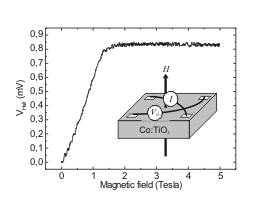}
\caption{\label{figure16} Hall effect data of sample 6 taken at 4.2
K. Inset shows the geometry of the Hall effect experiments. $H$ is
the external magnetic field applied perpendicular to the film
surface.}
\end{figure}

The Hall effect measurements were carried out at 4.2 K using a van
der Pauw configuration presented in Fig.~\ref{figure16} as an inset.
In spite of the fact that the structural and the magnetization
measurements indicate the presence of magnetic nanoparticles in the
Co-implanted TiO$_2$ films, the anomalous Hall effect is observed
for these samples. The Hall effect data of sample 6 are shown in
Fig.~\ref{figure16}. As it is explained above, a rapid increase in
the Hall voltage at low fields can be interpreted as an AHE which is
followed by a slow decrease corresponding to the ordinary Hall
effect. It is important to note that the negative slope of the high
field data indicates \emph{n}-type carriers in Co-implanted TiO$_2$
rutile. The electron density (n), calculated from the slope of the
curve at higher fields, is about $3.75\times10^{18}cm^{-3}$.

\section{Discussion}

The origin of the observed two magnetic phases (ferromagnetism and
superparamagnetism) in Co-implanted TiO$_2$ rutile plates is
attributed to the formation of two cobalt enriched layers with
different cobalt concentrations and valence states of the cobalt
dopant. The TEM images (Fig.~\ref{figure6}) clearly show that
nanosize magnetic particles of cobalt metal nucleate in the surface
region of the implanted rutile where the cobalt concentration is
maximal (see RBS data in Fig.~\ref{figure1}). Mostly beneath this
layer, in the tail of the depth profile, the implanted cobalt can
exist in an ionic state substituting the Ti$^{4+}$ ions in the
matrix by Co$^{2+}$ ions. Thus for charge neutrality either two
Co$^{2+}$ ions substitute for one Ti$^{4+}$ ion, or one Co$^{2+}$
ion and one oxygen vacancy are formed. At the lowest dose
($0.25\times10^{17} ions\cdot cm^{-2}$) the magnetic contribution
from the metallic cobalt clusters and the substituted cobalt ions is
very small, and hence the MOKE signal at room temperature is rather
weak. Increasing the cobalt implantation dose leads to both, an
increase of the Co cluster size, as seen by the increasing blocking
temperature, and to more substitutional cobalt in the Co$^{2+}$
state.  At certain concentrations the substituted cobalt ions start
to interact leading to ferromagnetism at room temperature in sample
4 ($1.00\times10^{17} ions\cdot cm^{-2}$) and sample 5
($1.25\times10^{17} ions\cdot cm^{-2}$). At doses higher than
$1.25\times10^{17} ions\cdot cm^{-2}$ strong ferromagnetic order is
formed due to the ion accumulation and indirect exchange interaction
between the Co$^{2+}$ ions. However, TEM images, peaks in the ZFC
curves, and two component hysteresis curves for dose levels
$1.25-1.50\times10^{17} ions\cdot cm^{-2}$ indicate that the
superparamagnetic phase is present in these samples and coexists
with the ferromagnetic phase.

Revealing the interaction mechanism of substituted cobalt ions which
leads to ferromagnetism in Co:TiO$_2$ is also an important incentive
of this study. Since the XAS spectra clearly show the multiplet
structure of the Co \emph{L}$_3$ peak (see Fig.~\ref{figure12}), it
is certain that some portion of the implanted cobalt ions in TiO$_2$
rutile are in the Co$^{2+}$ oxidation state. When TiO$_2$ is doped
with cobalt ions, simultaneously oxygen vacancies are also expected
to be produced \cite{ErricoPRB05}. The observation of the AHE in the
Co-implanted rutile samples give clear evidence for oxygen vacancies
which contribute to shallow donor states in TiO$_2$ and increase the
carrier density \cite{YahiaPR63}. It was suggested that these oxygen
vacancies strongly promote ferromagnetism in Co-implanted TiO$_2$
films by an indirect exchange of substituted cobalt ions through
electrons trapped by neighboring oxygen vacancies
\cite{CoeyNature05}.

We have also noticed a clear polarization of the oxygen
\emph{p}-orbitals in Co-implanted TiO$_2$ rutile samples. The shape
of the hysteresis curve and the coercive field measured at the O
\emph{K} edge is the same as the one recorded at the Co \emph{L}$_3$
edge \cite{NefedovAPL06}. This is a clear indication that oxygen
ions, which are close to Co ions in TiO$_2$ become magnetically
polarized. Whether the oxygen polarization is essential for
supporting ferromagnetic exchange is presently not clear.

Another important result of this study is the observation of an
anomalous Hall effect. The AHE is often taken as an evidence that
the charge carriers are polarized and that the material is a true
DMS. However, after simultaneous observation of superparamagnetism
and AHE in Co-doped TiO$_2$ films by Shinde \emph{et al.}
\cite{ShindePRL04} and also in this study, the existence of an AHE
can be thought of a necessary measurement condition but it is not
sufficient by itself to claim the intrinsic nature of ferromagnetism
in a DMS material.

\section{Summary and conclusions}

In conclusion, we have studied in detail the structural, magnetic
and electronic properties of Co-implanted TiO$_2$ rutile films for
different implantation doses. The structural data clearly show that
cobalt clusters are present in the samples after high dose cobalt
ion implantation. In addition to the cluster formation, substitution
of cobalt ions into the rutile lattice is also confirmed by XAS
experiments. The origin of the observed magnetic behavior in the
samples is explained by the coexistence of two different magnetic
phases. Cobalt nanoparticles in the surface layer form a
superparamagnetic phase in the samples implanted with low and
intermediate doses. In addition, substitution of Ti$^{4+}$ ions by
Co$^{2+}$ ions leads to intrinsic ferromagnetism as a second
magnetic phase. The oxygen vacancies formed by ion implantation
provide charge compensation and serve as mediators for the exchange
interaction between the Co$^{2+}$ ions in high dose doped samples.
The observation of the anomalous Hall effect in Co-implanted TiO$_2$
rutile can also be thought of an important indicator for the
observed long range ordered intrinsic ferromagnetism in the rutile
phase. At the highest dose, a strong ferromagnetic phase exists with
a Curie temperature of above 700 K. This ferromagnetic phase
exhibits a perfect uniaxial in-plane magnetic anisotropy following
exactly the crystal symmetry of the TiO$_2$ rutile. We consider this
as the strongest experimental evidence for the intrinsic nature of
ferromagnetism in the Co-doped TiO$_2$ rutile.

\subsection{Acknowledgments}
We wish to acknowledge A. Kr\"{o}ger for preparation of TEM samples,
and O. Seeck and W. Caliebe (HASYLAB) for their assistance with the
beamline operation. This work was partially supported by BMBF
through Contracts Nos. 05KS4PCA (ALICE Chamber) and 05ES3XBA/5
(Travel to BESSY), by DFG through SFB 491, and by RFBR through the
grant Nos 07-02-00559-a and 04-02-97505-r. N. Akdogan acknowledges a
fellowship through the IMPRS-SurMat.

\section*{References}
\bibliography{JPDAP_Akdogan}

\end{document}